\documentstyle[preprint,prd,aps,psfig]{revtex}
\tightenlines

\newcommand{\beq}{\begin{equation}}
\newcommand{\eeq}{\end{equation}}
\newcommand{\beqa}{\begin{eqnarray}}
\newcommand{\eeqa}{\end{eqnarray}}
\newcommand{\simg}{\gtrsim}
\newcommand{\siml}{\lesssim}

\begin{document}

\draft
\preprint{hep-ph/9905401, UTAP-326, KNGU-INFO-PH-8}

\title{
Formation of multiple winding topological defects \\
in the early universe
}

\author{Takahiro Okabe\footnote{
Electronic address: okabe@utaphp1.phys.s.u-tokyo.ac.jp
}
}

\address{
Department of Physics, University of Tokyo, 
Tokyo 113-0033, Japan
}

\author{Michiyasu Nagasawa\footnote{
Electronic address: nagasawa@info.kanagawa-u.ac.jp
}
}

\address{
Department of Information Science, Faculty of Science,Kanagawa University, \\
Kanagawa 259-1293, Japan
}

\maketitle

\begin{abstract}

The formation probability of multiple winding topological defects is
calculated by phase and flux distribution analysis based on the Kibble
mechanism.
The core size of defects is taken into account so that when it is much
larger than the correlation length of the Higgs field, high winding
configuration can be realized.
For example, if the self coupling constant of the Higgs field is
smaller than $4\times10^{-4}$, the topological inflation may occur at
the grand unification energy scale by the multiple winding string
produced during the GUT phase transition.

\end{abstract}

\pacs{PACS numbers: 98.80.Cq, 11.27.+d, 11.30.Fs}

\section{introduction}

It is believed that topological defects are produced during certain types
of cosmological phase transitions and may play an important role in
the course of cosmic evolution\cite{k76,vs94}.
They can solve some of the unsolved cosmological problems.
Particularly topological defects give a natural and attractive model
for inflation and may explain the source of baryon asymmetry in our universe.

Inflation is the most promising candidate which can provide the
solutions to the problems in the standard Big Bang
theory\cite{inflation} and the realistic model for the inflation is
very important subject in the cosmology.
Topological inflation has been proposed by Vilenkin\cite{v94} and
Linde\cite{l94}.
In this inflation scenario, the energy density which drives the
inflationary expansion of the universe is provided by the symmetric
state within the defect core where the false vacuum energy is trapped
by the topological constraint.
If the way of the symmetry breaking in the particle physics theory
satisfies the condition for the production of the topological defect, the
formation of some kind of defect at the phase transition is inevitable.
Then one of the necessary conditions for the inflation that the
inflaton field must be in the state which has sufficient vacuum energy
density would be realized without any additional constraint.
This is the advantage of the topological inflation scenario since in
the conventional inflation model, the fine-tuning of initial state
selection is required in order that the necessary conditions for
the inflation should be satisfied\cite{l90}.
However, in order to realize the condition that the core length scale
is long enough to cover the horizon scale, the symmetry breaking
energy scale for the defect formation, $\eta$, should be larger than
the Planck scale as
\beq
\eta \simg M_{pl}\ .
\eeq
Therefore in order to understand the topological inflation in detail,
we must work very close to the Planck scale at which our classical field
theories will not be valid.
When multiple winding defects are employed, however, the energy scale
at which the inflation occurs is decreased and the constraint on $\eta$
is relaxed since the higher winding defect has thicker core length
scale than the unit winding defect.
The constraint on $\eta$ when the winding number of the string, $n$,
is included has been derived numerically \cite{d98} as
\beq
\eta \simg 0.16M_{pl}\times n^{-0.56}\ .\label{eq:topinf}
\eeq
Hence if the string whose winding number $n \sim 3\times10^3$ is
produced, it is possible that the topological inflation occurs at the
GUT scale.

The baryon asymmetry problem is another important subject.
Since the sphaleron transition interaction should erase the baryon
asymmetry produced before such a process becomes negligible unless
the difference between lepton number and baryon number exists,
the baryon number generation at the electroweak scale seems to be
the most conventional scenario of the baryogenesis at present.
The first proposed electroweak baryogenesis scenario relies on having
a strong first order phase transition so that the deviation from
the thermal equilibrium state, one of the necessary conditions for
the baryogenesis, is achieved by the propagation of nucleated bubbles
in the plasma\cite{krs85}.
However, it seems that the first order electroweak phase transition is
difficult to be established in the standard model\cite{t98}.
Then the electroweak baryogenesis scenario using electroweak strings
was proposed by Brandenberger and Davis\cite{bd93}.
In this scenario the out-of-equilibrium condition is provided by the
collapse of strings.
Thus the string baryogenesis has the advantage that it works effectively
whether the electroweak phase transition is of first order or not.
Since the electroweak string in the standard model is topologically
unstable, however, its formation probability is too small to serve for
the observational amount of baryon number\cite{ny96}.
Therefore topological defects associated with the electroweak symmetry
breaking should be necessary for the electroweak baryogenesis.
Recently Soni\cite{s97} has suggested a new scenario for the baryon
asymmetry generation using the string produced at the electroweak
energy scale.
He has pointed out that the sphaleron energy in the presence of the
string with a few winding number can become negative.
Then when the strings within which sphalerons are bounded decay, the
baryon number would be produced.
In this scenario the thermal equilibrium is violated by the
string-sphaleron system which is left below the sphaleron suppression
temperature.

In order to know to how much extent the required energy scale for $\eta$
becomes small in the topological inflation scenario, we have to estimate
the possibility of multiple winding defect formation.
Also in order to determine how much baryon asymmetry is produced in
the electroweak baryogenesis scenario by the sphaleron bound state on
the string, we have to calculate the formation probability of multiple 
winding strings.
For these reasons, in this Letter we consider the realization of
multiple winding topological defect configuration.
We employ two kinds of defect, that is, strings and monopoles.
For the string case, we investigate the phase distribution of the Higgs
field.
For the monopole, the distribution of the gauge flux is considered.
In both cases, the formation probability of multiple winding defects
is estimated quantitatively and the existence of such a defect is
confirmed.

\section{string}

First let us consider the breaking of a local $U(1)$-symmetry in the
abelian-Higgs model with Lagrangian:
\beq
{\cal L} = - \frac{1}{4}F^{\mu\nu}F_{\mu\nu}
     - \frac{1}{2}(D^{\mu}\phi)^{\dagger}(D_{\mu}\phi)
     - \frac{1}{8}\lambda({\phi}^{\dagger}\phi - {\eta}^2)^2\ ,
\eeq
where $\phi$ is a complex scalar field and the covariant derivative is
given by $D_{\mu} = {\partial}_{\mu} - ieA_{\mu}$ with $A_{\mu}$
a gauge vector field, $e$ the gauge coupling constant.
$F_{\mu\nu}$ is the antisymmetric tensor defined by
$F_{\mu\nu} = {\partial}_{\mu}A_{\nu}-{\partial}_{\nu}A_{\mu}$.
It is well known that there is a string solution called the
Nielsen-Olesen vortex line\cite{no73} in this model.
In the Lorentz gauge, the Higgs field configuration far from the core
of the string can be written as
\beq
\phi \simeq \eta e^{in\theta}\ .
\eeq
The winding number is a strictly conserved quantity and the total Higgs
field phase difference around the string is $2\pi n$.
In general, the larger $n$ becomes, the line energy density of the string
increases and the core width scale becomes fatter.
It has been demonstrated both numerically and analytically that
multiple winding strings $(|n| > 1)$ are stable when $\lambda / e^2 <
1$ and unstable in the opposite case\cite{stability}.
In the former case, a number of string may get together and coalesce
into one string due to the energetic favorableness.
On the other hand, in the latter case, multiple winding strings can
break up into $|n|$ pieces of string with unit winding number.

Before estimating the formation probability of multiple winding
strings, let us briefly summarize the estimation method for the unit
winding string based on the conventional Kibble mechanism.
In the thermal phase transition, the field has a typical length scale,
that is, the correlation scale, $\xi$.
It defines the size of regions within which the values of fields are
homogeneous and are independent of those at other regions.
When the cosmic temperature decreases sufficiently and the ground state
of the Higgs field becomes the true vacuum state, the amplitude of
the Higgs field $|\phi|$ should be same almost everywhere, while its
phase $\theta$ varies on the correlation scale.
Then the physical space is regarded to be divided to correlated
volumes and the phase of the Higgs field takes the random value at
each correlated region so that the phase distribution has a
domain-like structure at the end of cosmological phase transition.
Thus in the context of the Kibble mechanism, the formation probability
of the string can be estimated as follows\cite{vv84,p91}.
For simplicity, the 2-dimensional slice of the 3-dimensional $\theta$
distribution is considered so that the formation of vortices can be
analyzed instead of strings.
\begin{enumerate}
\item[(A)]Divide the plane into 2-dimensional domains whose typical size
is equal to the correlation length of the Higgs field $\xi$.
\item[(B)]Assign the phase of the Higgs field randomly to one representative
point of each domain.
\item[(C)]Interpolate the phase between two representative points of
neighboring domains so that the gradient energy of the Higgs field takes
minimum value (geodesic rule).
\end{enumerate}
Then we can count the total phase change along any closed loop on
the plane and calculate how much winding number exists inside the loop
so that the formation probability of the vortex can be estimated.
Usually it is assumed that at most three different domains meet at the
boundary point.
It means the closed loop which is used to count the phase change can
always be identified to the triangle whose corners correspond to three
representative points of these three domains.
Thus the above procedure (A) and (B) can be expressed as: divide the
plane by regular triangles and assign the phase of the Higgs field
randomly to each vertex point where six triangles join.
The geodesic rule implies that the difference of the phase between two
neighboring points is less than $\pi$.
Therefore when the triangle division is imposed, the total difference
of the phase along the circumferential edge of the triangle is $2\pi$
at most because of the phase continuity.
As a result, the winding number should not exceed the unity and
multiple winding vortices never appear in this situation.

\begin{figure}

\centerline{\psfig{file=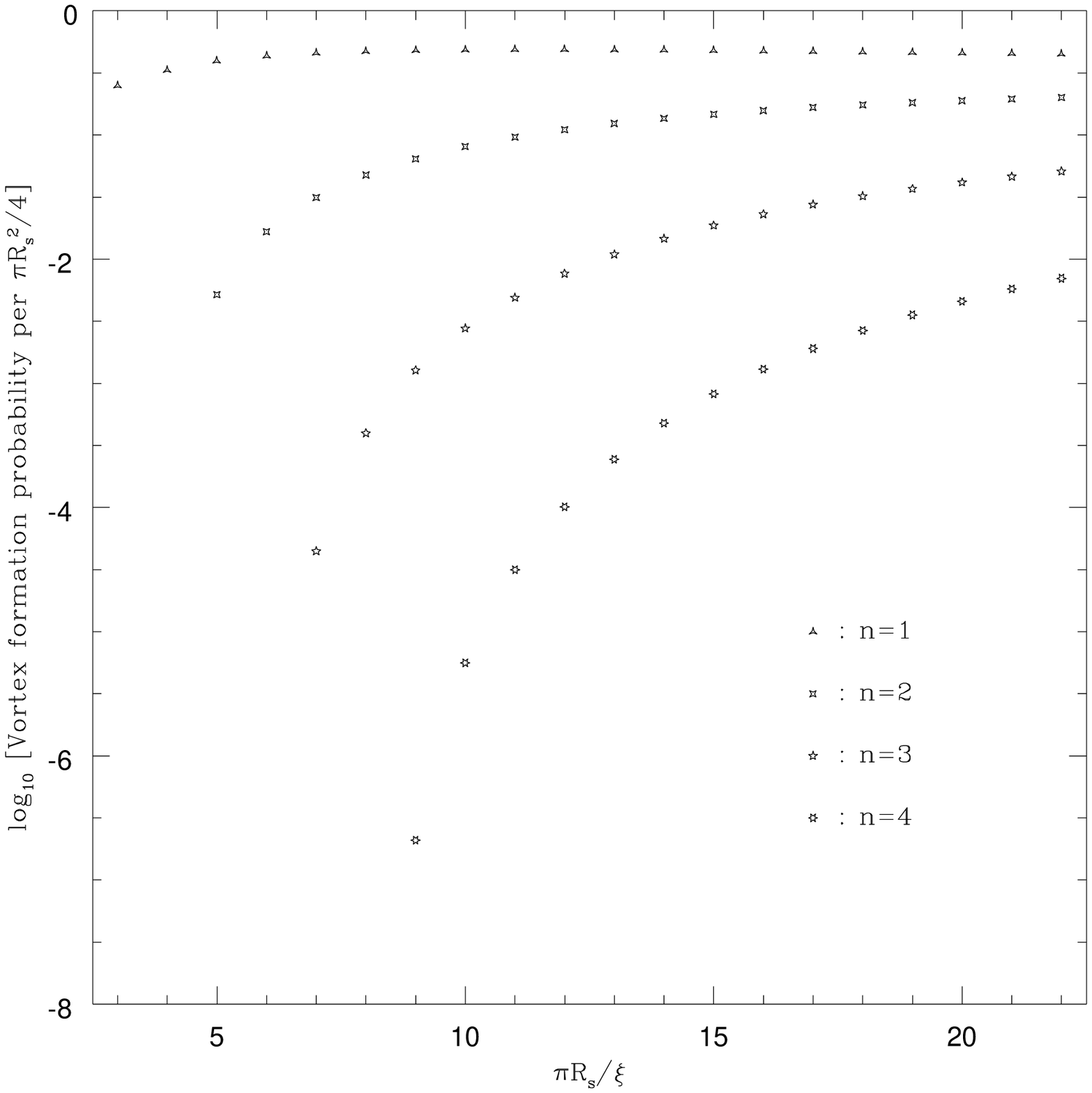,width=\columnwidth}}
\caption{Formation probability of vortices per one polygon for various
values of $\pi R_s/\xi$. $n$ depicts the winding number.}
\label{fig:string}

\centerline{\psfig{file=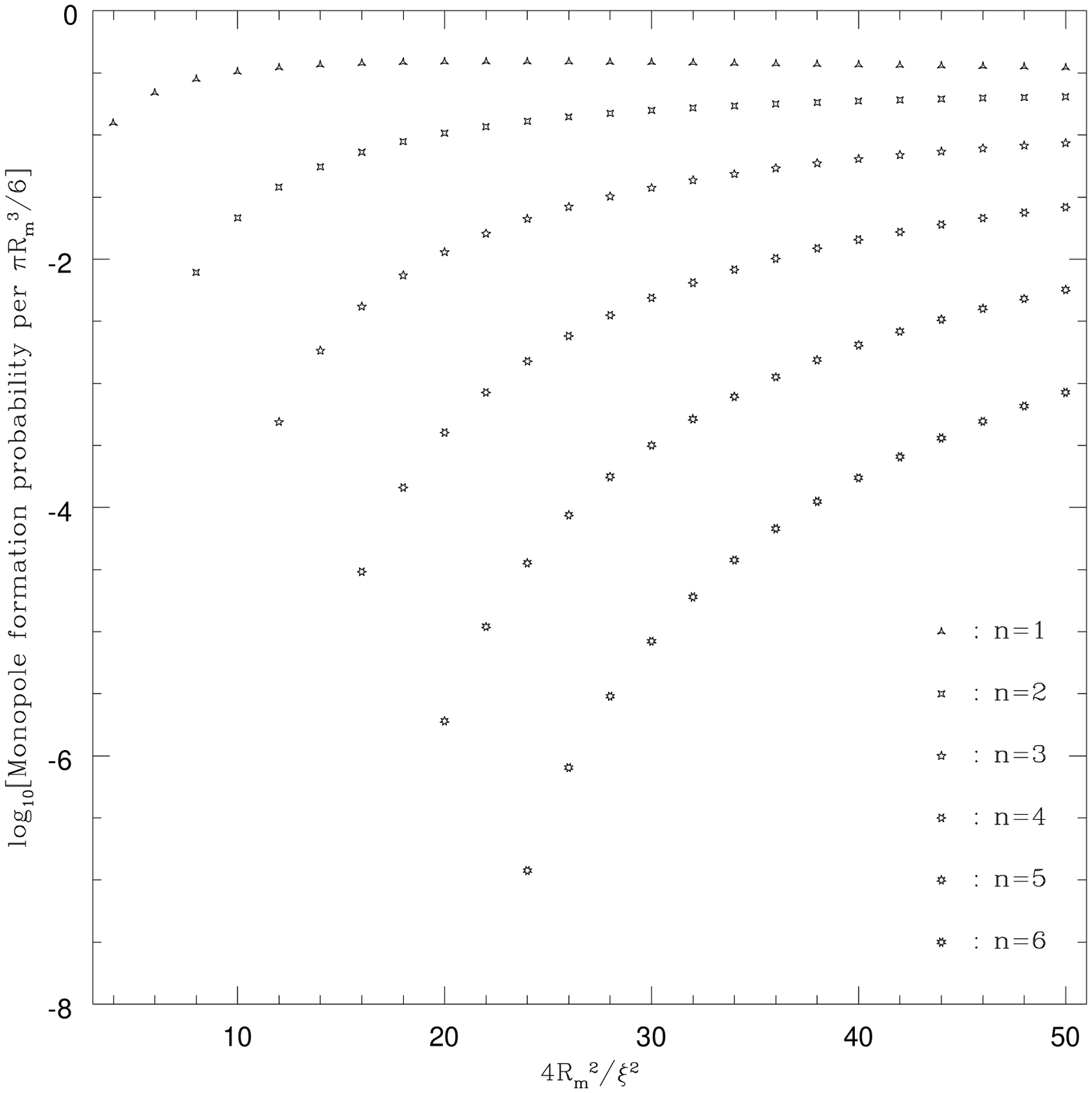,width=\columnwidth}}
\caption{Formation probability of monopoles per one domain for various
values of $4{R_m}^2/\xi^2$. $n$ depicts the winding number.}
\label{fig:monopole}

\end{figure}

However, we cannot say that formation probability of multiple winding
vortices equals zero since this estimation is based on too simplified
assumptions.
In order to construct the method which can detect the existence of
multiple winding number, we modify the arrangement of points where the
phase of the Higgs field is allocated.
Since so long as the plane is divided into triangles the maximum
winding number must be only one, the plane should be divided by a
polygon other than a triangle so that the total phase difference along
its periphery can go beyond $2\pi$.
It would be reasonable if we choose the side length of the polygon as
$\xi$ since each vertex can be considered to represent the correlated
domain in the similar manner to the triangle case.
When the diameter scale of the polygon is $R_s$, the total length of
the polygon periphery will be $\sim \pi R_s$.
Then the number of vertices is $\sim \pi R_s/\xi$ and the possible
largest phase change is $\sim \pi^2 R_s/\xi$.
In the usual triangle case, $\pi R_s/\xi =3$ so that $R_s\sim \xi$.
The revised winding number counting procedure can be expressed as:
\begin{enumerate}
\item[(A')]Divide the plane into polygons whose diameter scale is
  equal to $R_s$ and vertex number is $\pi R_s/\xi$.
\item[(B')]Assign the phase of the Higgs field randomly to each vertex 
  of the polygon.
\end{enumerate}
The third step is identical to the original version.
Then we can count the winding number and estimate the formation
probability of multiple winding vortices.
In actual calculations, we consider one polygon, assign a phase of the 
Higgs field ($0\le \theta <2\pi$) randomly to each vertex of this
polygon and count the winding number along its periphery.
We repeat this process $10^8$ times so that the probability
distribution of  the string formation for each winding number can be
led.
The result of calculations for various values of $\pi R_s/\xi$ is
shown in FIG.\ \ref{fig:string}.

In our method, the numerical value of the multiple winding vortex
formation probability depends on $R_s/\xi$.
It would be reasonable that $\xi$ can be obtained by the correlation
scale of the Higgs field at Ginzburg temperature $T_G$, when the phase
transition terminates and the defect cannot be erased by thermal
fluctuations, as
\beq
\xi \sim \frac{1}{T_G}\ ,
\eeq
because the correlation length of the massless field corresponding to
thermal fluctuations is given by the inverse of temperature\cite{yy97}.
The most natural estimation of $R_s$ would be that it is comparable to
the core diameter of the string.
Under this assumption, the number of string which can be produced
inside each polygon should be one at most since many strings cannot
exist within the scale of the string core or in other words all the
winding number one polygon contains must belong to a single string.
Thus we do not mistake two or more strings for one string if we take
$R_s$ to be the core diameter of the string.
The string core diameter scale is given by the Compton wavelength of
the Higgs field as
\beq
R_s = \frac{1}{m_H(T_G)} \sim \frac{1}{\lambda T_G}\ ,\label{eq:rs}
\eeq
at $T_G$ where $m_H$ is the mass of the Higgs field.
Therefore the maximum winding number the string can have will be
\beq
n_{max} \sim \Bigg[\frac{\pi^2 R_s}{2\pi\xi}\Bigg] \sim
\Bigg[\frac{\pi}{2\lambda}\Bigg]\ , \label{eq:ratios}
\eeq
where we use the Gauss's symbol.
This result means multiple winding string can be produced when
$\lambda$ is smaller than $\pi /4$.

When $\lambda \sim 1$, the formation of multiple winding strings
seems to be impossible even in the revised method.
There is, however, another possibility that the correlation length of
the Higgs field can fluctuate since in reality the size of the domain
within which the value of the field is homogeneous may not be constant
throughout the universe.
As a toy model, we assume that the domain size distributes around its
averaged value $\xi$ and the probability distribution function obeys
a Gaussian form whose dispersion is given by $\sigma$.
Then we calculate the distribution function of $\pi R_s/\xi,
P(\pi R_s/\xi)$, and estimate the formation probability of the string
for each winding number by summing up that for fixed value of
$\pi R_s/\xi$ shown in FIG.\ \ref{fig:string} with weight $P(\pi R_s/\xi)$.
Here, we set $\xi = \sigma = R_s$.
Note that the larger $\sigma$ becomes, larger domains which contains no
string tend to be produced.
The resulting formation probability, $P_s(n)$, of the vortex for $n=1-4$
per one domain is shown in TABLE \ref{tab:string}.
The larger $n$ becomes, the formation probability of multiple winding
vortices decreases exponentially.

\section{monopole}

The formation probability of multiple winding strings depends on the
ratio of the circumference of the string to the correlation length of
the Higgs field.
Also in the case of monopole, a similar consideration can be applied.
The formation probability of multiple winding monopoles depends on the
ratio of the surface area of monopole to the correlation area of the
Higgs field, which is given by $4\pi {R_m}^2/\pi{\xi}^2$, where $R_m$
is the core diameter of the monopole.
Therefore in principle we can calculate the formation probability of
multiple winding monopoles for various values of $4 {R_m}^2/\xi^2$
in a similar manner for strings.
However, since it is complicated to interpolate the phase of the Higgs
field between two neighboring vertices and count the winding number,
here we resort to an easier method.

In the case of local monopole, not only the Higgs field but also the
gauge field appears in the theory.
Then instead of the phase of the Higgs field, we assign the magnetic
flux of the gauge field at each domain.
When the total flux which passes through a closed surface is not zero,
there must be a monopole or an anti-monopole inside this surface.
Thus the gauge flux can play a role similar to the winding number.
The simplest monopole solution, the 't Hooft-Polyakov
solution\cite{tp} for the model in which the $SO(3)$ symmetry is
broken to $U(1)$, has the overall magnetic flux whose magnitude is
\begin{equation}
{\Phi}_B = \frac{4\pi}{e}\ ,
\end{equation}
which corresponds to the unit winding number in the Higgs field case.
The 't Hooft-Polyakov monopole with multiple winding number is known
to be stable in the Prasad-Sommerfield limit\cite{ps}.
In actual calculations, we assign only discrete values of magnetic flux
as $\pm \pi/e$ for simplicity.
Then the total amount of the flux is summed up around the closed
surface and the winding number can be calculated.
Then the formation probability of the multiple winding monopole is given by
\beq
P_m(n,l) = \left( {}_l C_{\frac{l}{2}+2n} + {}_l C_{\frac{l}{2}+2n+1} \right)
\left(\frac{1}{2}\right)^{l-1}\ ,\label{eq:mwm}
\eeq
where $n$ is the winding number and $l = 4{R_m}^2/\xi^2$.
We omit the fractional part of the winding number which appears because of
the absence of the interpolation process.
Since our method of the magnetic flux assignment reproduces the formation
probability of a unit winding monopole\cite{p91} when $l=4$, the
approximation we have employed should be reasonable.
The analytic result for various values of $4{R_m}^2/\xi^2$ using
the equation (\ref{eq:mwm}) is shown in FIG. \ref{fig:monopole}.

Similarly to the string case, $R_m$ can be estimated to be
$(\lambda T_G)^{-1}$ and $\xi$ equals $T_G^{-1}$.
Then the ratio of the surface area of the monopole to the correlation
area of the Higgs field is given by
\beq
\frac{4{R_m}^2}{\xi^2} \sim \frac{4}{\lambda^2}\ .
\eeq
Also in this case the smaller $\lambda$ becomes, the more multiple
winding monopoles are produced.
Moreover, the spatial variation of the correlation length enables the
realization of the high winding configuration in the same way for the
string.
The resulting formation probability, $P_m(n)$, of the monopole for
$n=1-4$ per one domain is shown in TABLE \ref{tab:monopole}.

\section{conclusion}

In the present Letter we have calculated the formation probability of
multiple winding topological defects by means of the Kibble mechanism.
Since we have taken the fact that the core size scale of the defect
can be larger than the correlation scale of fields into account, the
formation probabilities depend on the ratio of these two scales.
The multiple winding topological defects can be produced when the self
coupling constant of the Higgs field, $\lambda$, which defines the
defect core thickness is much less than the unity.
The smaller $\lambda$ becomes, the formation probability of the
multiple winding defects and the maximum winding number the defect can 
have increase.

This result gives following cosmological implications.
First it may be possible that the topological inflation occurs at the
GUT energy scale by the multiple winding string produced during the
GUT phase transition when $\lambda \siml 4\times10^{-4}$ which can be
calculated from the equations (\ref{eq:topinf}) and (\ref{eq:ratios}).
For the same value of $\lambda$, the maximum winding number the
monopole can have is larger than that the string can have.
Therefore in the case of monopole, topological inflation may occur for 
larger value of $\lambda$.
Secondly, in the context of the electroweak baryogenesis, the
sphaleron bound state scenario using the string whose winding number
is $2\sim 3$ will be promising.
Since the formation probability of the double winding string is about
one third that of a unit winding string when $\lambda \sim 10^{-1}$, it
might be possible to explain the observational amount of the baryon
asymmetry.
Further quantitative analysis will be needed.

Our method is not the only one which can improve the simple estimation
of the defect formation probability.
For example, not only the correlation length scale of the Higgs filed
but also the core thickness scale of the topological defect would vary
since the static and highly symmetric solution of the defect
configuration may not always be applied to the actual situation.
Another promising modification is the relaxation of the geodesic
rule\cite{pv98}.
It is too simplified picture that the value of the field should be
interpolated through the shortest path on the vacuum manifold and
there is a possibility that larger gradient exists between two regions.

In addition to such revisions, the interaction between strings makes
the situation extremely better.
We have considered only the moment of string production and neglected
the dynamics of the string after its formation when we have calculated
the value of $R_s$.
In some ranges of the parameter, however, the attractive force
operates on strings so that the winding number can accumulate.
It corresponds to the case when $R_s/\xi$ is enormous.
Maybe a single string has all the winding number within the horizon
scale.
Then the topological inflation scenario might successfully work even
when $\lambda$ is not so small.
Note that also the interaction of the string with the surrounding
plasma will affect the dynamical evolution.
The drag force can aid the accumulation process because it dissipates
the energy of the string pair and allows a bound state to be stable.

\acknowledgments

T.O. is grateful to Professor Katsuhiko Sato and Professor Yasushi
Suto for their encouragement and to Masahide Yamaguchi for discussion.
M.N. thanks Leandros Perivolaropoulos for his comment at the workshop
in Les Houches.

\begin{center}

\begin{table}
 \caption {Formation probability of vortices per one polygon. The
  fluctuation of the correlation length is taken into account.}
 \label{tab:string}
\end{table}

\begin{tabular}{lr}\hline\hline
   $\ \ n\ \ $ & $\ \ \ \ \ \ \ P_s(n)\ \ \ \ \ \ \ $ \\ \hline
   $\ \ 1\ \ $ & $2.102 \times 10^{-1\ }\ $ \\
   $\ \ 2\ \ $ & $8.36 \times 10^{-4\ }\ $ \\
   $\ \ 3\ \ $ & $4.8 \times 10^{-7\ }\ $ \\
   $\ \ 4\ \ $ & $8 \times 10^{-11}\ $ \\
   \hline\hline
\end{tabular}

\vspace{1cm}

\begin{table}
 \caption {Formation probability of monopoles per one polyhedron. The
  fluctuation of the correlation length is taken into account. Here,
  we set $\xi = \sigma = R_m$.}
 \label{tab:monopole}
\end{table}

\begin{tabular}{lr}\hline\hline
   $\ \ n\ \ $ & $\ \ \ \ \ \ \ P_m(n)\ \ \ \ \ \ \ $ \\ \hline
   $\ \ 1\ \ $ & $4.285 \times 10^{-2\ }\ $ \\
   $\ \ 2\ \ $ & $7.36 \times 10^{-5\ }\ $ \\
   $\ \ 3\ \ $ & $3.4 \times 10^{-8\ }\ $ \\
   $\ \ 4\ \ $ & $7 \times 10^{-12}\ $ \\
   \hline\hline
\end{tabular}

\end{center}

\end{document}